\documentclass[showpacs,preprintnumbers,amsmath,amssymb]{revtex4}
\newtheorem{theorem}{Theorem}
\newtheorem{definition}{Definition}

\newtheorem{corollary}{Corollary}

\usepackage{graphicx}
\usepackage{dcolumn}
\usepackage{bm}

\begin{document}
\title{
Quantum Kolmogorov Complexity and Information-Disturbance Theorem
}

\author{Takayuki Miyadera}
\affiliation{%
Research Center for Information Security (RCIS), \\
National Institute of Advanced Industrial
Science and Technology (AIST). \\
Daibiru building 1003,
Sotokanda, Chiyoda-ku, Tokyo, 101-0021, Japan.
\\
(E-mail: miyadera-takayuki@aist.go.jp)
}%


\date{\today}

\begin{abstract}
In this paper, a representation of the 
information-disturbance theorem 
based on the quantum Kolmogorov complexity that was defined by 
P. Vit\'anyi has been examined. 
In the quantum information theory, 
the information-disturbance relationship, which 
treats the trade-off relationship between information gain and 
its caused disturbance,  
 is a fundamental result that is related to 
 Heisenberg's uncertainty principle. 
The problem was formulated in a cryptographic setting 
and quantitative relationships between 
complexities have been derived. 
\end{abstract}
\maketitle

\section{Introduction}
The quantum theory enables us to process information in ways 
that are not feasible in the classical world. 
Quantum computers can solve difficult problems 
such as factoring \cite{ShorFactor} 
or searching \cite{Grover} 
in drastically small time steps. 
Quantum key distribution \cite{BB84,E91}
achieves information-theoretic security 
unconditionally. 
This field of the quantum 
information theory has been intensively 
studied during the last two decades. 
While 
most of the studies in this field
investigate
how Shannon's information theory was modified or
restricted by the quantum theory, 
there is another information theory 
called the algorithmic information theory
\cite{Kolmogorov, Chaitin}.
In contrast to 
Shannon's theory, which defines 
information 
using a probability distribution, 
the algorithmic information theory
assigns the concept of information to 
individual objects by using a 
computation theory. 
Although the algorithmic information theory 
has been successfully applied to various fields 
\cite{LiVitanyi}, 
its quantum versions were only recently proposed
\cite{Vitanyi,Svozil,Gacs,vanDam}. 
We believe there have only been a few applications so far 
\cite{Mora,Benatti,MiyaQKD}.    
In this research, we study how
quantum Kolmogorov complexity, which was defined by Vit\'anyi,  
can be applied to demonstrate quantum effects 
in a primitive information-theoretic operation.
\par
We study the algorithmic information-theoretic representation 
of an information-disturbance relationship 
\cite{Boykin2,MiyaInfoDis,Miyaprep}, 
which addresses a
fundamental observation 
that information gain destroys quantum states.  
In particular,  
an operation that yields 
information gain with respect to 
an observable spoils quantum states that were 
prepared with respect to its conjugate (noncommutative) observable.
This relationship indicates the impossibility of 
jointly measuring noncommutative observables, and is therefore
it is related to 
Heisenberg's uncertainty principle. 
In addition, it plays a crucial role 
in quantum cryptography. 
Because a state is inevitably spoiled 
when an eavesdropper obtains information, 
legitimate users can notice the 
existence of the eavesdropper
\cite{Boykin}. 
In this study, we formulate the problem 
in a cryptographic setting and derive quantitative relationships. 
Our theorem, characterizing both the information gain 
and the disturbance in terms of the quantum Kolmogorov 
complexity, demonstrates a trade-off relationship between these 
complexities. 
\par
This paper is organized as follows. 
In the next section, we give a brief review of 
quantum Kolmogorov complexity defined by Vit\'anyi.
In section \ref{mainsection}, we introduce 
a toy quantum cryptographic model and describe our main result
on the basis of this model. 
The paper ends a short discussion. 
\section{
Quantum Kolmogorov Complexity based on Classical Description }
\label{sec:QKC}
Recently some quantum versions of Kolmogorov 
complexity were proposed by a several researchers.   
Svozil \cite{Svozil}, in his pioneering work, 
defined the quantum Kolmogorov complexity 
as 
the minimum classical description length of a quantum state
through a quantum Turing machine \cite{Deutsch,Bernstein}. 
As is easily seen by comparing the cardinality of a set of all the 
programs with that of a set of all the quantum states, 
the value often becomes infinity. 
Vit\'anyi's definition \cite{Vitanyi}, 
while similar to Svozil's, does not 
have this disadvantage. 
Vit\'anyi added a term that compensates for 
the difference between a target state and an output state. 
Berthiaume, van Dam and Laplante \cite{vanDam}
defined 
their quantum Kolmogorov complexity as the length of the shortest 
quantum program that outputs a target state.
The definition was settled and its properties 
were extensively investigated by M\"uller \cite{Muller, MullerPhD}.
Gacs \cite{Gacs} employed a different 
starting point related to the algorithmic probability
to define his quantum Kolmogorov complexity.     
\par
In this paper we employ a definition given by 
Vit\'anyi \cite{Vitanyi}.
His definition based on the classical description length is 
suitable for quantum information-theoretic problems which 
normally treat classical inputs and outputs.   
In order to explain the definition precisely, 
a description of one-way quantum Turing machine 
is needed. 
It is utilized to define a prefix quantum Kolmogorov complexity. 
A one-way quantum Turing machine consists of 
four tapes and an internal control. 
(See \cite{Vitanyi} for more details.)
Each tape is a one-way infinite qubit (quantum bit) chain
and has a corresponding head on it. 
One of the tapes works as the input tape and 
is read-only from left-to-right.
A program is given on this tape
as an initial condition.
The second tape works as the work tape.
The work tape is initially set to be $0$ for 
all the cells. The head on it can read and write 
a cell and can move in both directions.
The third tape is called an auxiliary tape.
One can put an additional input on this tape. 
The additional input is written to the leftmost qubits and 
can be a quantum state or a classical state. 
This input is needed when one treats conditional 
Kolmogorov complexity. 
The fourth tape works as the output tape. 
It is assumed that after halting
 the state over this tape will not be changed.
The internal control is a quantum system 
described by a finite dimensional 
Hilbert space 
which has two special orthogonal vectors
$|q_0\rangle$ (initial state) and $|q_f\rangle$ (halting state). 
After each step one makes a measurement of a coarse grained 
observable on the internal control $\{|q_f\rangle \langle q_f|,
{\bf 1} -|q_f\rangle \langle q_f|\}$ to know if 
the computation halts. 
Although there are subtle problems 
\cite{Myers, Ozawa, Popescu, Miya}
in the halting process of 
the quantum Turing machine, we do not get into this problem and 
employ a simple definition of the halting. 
A computation halts at time $t$ if and only  
if 
the probability to observe $q_f$ at time $t$ is $1$, 
and at any time $t'<t$ the probability to observe 
$q_f$ is zero.  
By using this one-way quantum Turing machine,  
Vit\'anyi defined the quantum Kolmogorov complexity as 
the length of the shortest description of a quantum state.
That is, the programs of quantum Turing machine are restricted to 
classical ones, while 
the auxiliary inputs can be quantum states. 
We write $U(p,y)=|x\rangle$ if and only if 
a quantum Turing machine $U$ with 
a classical program $p$ and an auxiliary 
(classical or quantum) input $y$ halts 
and outputs $|x\rangle$. 
The following is the precise description of 
Vit\'anyi's definition.
\begin{definition}\cite{Vitanyi}
The (self-delimiting) quantum Kolmogorov 
complexity of a pure state $|x\rangle$ 
with respect to a one-way quantum Turing machine $U$ with $y$ 
(possibly a quantum state) as conditional input 
given for free is 
\begin{eqnarray*}
K_U(|x\rangle|\ y)
:=\min_{p,|z\rangle} \{l(p)+ \lceil -\log|\langle z|x\rangle |^2 \rceil :
U(p,y)=|z\rangle\},
\end{eqnarray*}
where $l(p)$ is the length of a classical program $p$, and 
$\lceil a \rceil$ is the smallest integer larger than $a$. 
\end{definition}
The one-wayness of the quantum Turing machine ensures that 
the halting programs compose a prefix free set. 
Because of this, the length $l(p)$ is defined consistently. 
The term $\lceil -\log|\langle z|x\rangle |^2 \rceil$
represents how insufficiently an output $|z\rangle$ approximates 
the desired output $|x\rangle$. 
This additional term has 
a natural interpretation using the Shannon-Fano code. 
Vit\'anyi has shown the following invariance theorem, 
which is very important.  
\begin{theorem}\cite{Vitanyi}
There is a universal quantum Turing 
machine $U$, such that for all machines $Q$, 
there is a constant $c_Q$, such that for all quantum states $|x\rangle$
and all auxiliary inputs $y$ we have:
\begin{eqnarray*}
K_U(|x\rangle  |\ y)
\leq K_Q(|x\rangle |\ y)
+c_Q.
\end{eqnarray*}
\end{theorem}
Thus the value of 
quantum Kolmogorov complexity does not depend on 
the choice of a quantum Turing machine if 
one neglects the unimportant constant term $c_Q$. 
Thanks to this theorem, 
one often writes $K$ instead of 
$K_U$. 
Moreover, the following theorem is crucial for our discussion.
\begin{theorem}\label{th:classical}\cite{Vitanyi}
On classical objects (that is, finite binary strings that are all directly 
computable) the quantum Kolmogorov complexity 
coincides up to a fixed additional constant with the self-delimiting 
Kolmogorov complexity. That is, 
there exists a constant $c$ such that for 
any classical binary sequence $|x\rangle$,
\begin{eqnarray*}
\min_q \{l(q): U(q,y)=|x\rangle \}\geq
 K(|x\rangle|\ y)\geq 
\min_q \{l(q): U(q,y)=|x\rangle\} -c
\end{eqnarray*}
holds.
\end{theorem}
According to this theorem, for classical objects 
it essentially suffices to treat only programs that exactly 
output the object.
\section{Information-Disturbance trade-off}\label{mainsection}
In this section, we treat a toy model of quantum key distribution 
in order to discuss the information-disturbance relationship. 
Let us first review a standard scenario of 
quantum key distribution called BB84. 
Suppose that there exist three players
Alice, Bob, and Eve. 
Alice and Bob are legitimate users. 
Alice encodes a message in qubits 
with one of the bases $X$ or $Z$, 
and 
sends them to Bob. After confirming the
receipt of the qubits by Bob, she 
announces the basis that was used by her for encoding. 
If there is no eavesdropper, Bob can perfectly 
recover the message by simply measuring 
the qubits by using the disclosed basis. 
Conversely, if there exists an eavesdropper Eve, 
the state received by Bob is destroyed 
and he will be unable to recover the message in that case. 
More precisely, according to the information-disturbance 
theorem in Shannon's information-theoretical 
representation, Bob's state is inevitably 
spoiled 
when Eve employs an attack 
that helps her obtain information about the messages encoded 
in the conjugate basis. 
In order to accomplish the key distribution protocol, 
Alice and Bob perform an error correction 
followed by a privacy amplification. 
%
\par 
Motivated by this protocol, we introduce its toy version 
in order to investigate 
a universal relationship between information gain and disturbance. 
There are three players Alice, Bob and Eve. 
Alice chooses an
$N$-bit message $y\in \{0,1\}^N$
and a basis $X$ or $Z$ for its encoding. 
We write the standard basis of 
a qubit as $\{|0\rangle, |1\rangle\}$, 
which are eigenstates of $Z$. 
Its conjugate basis is written as 
$\{|\overline{0}\rangle, |\overline{1}\rangle\}$, 
which are eigenstates of $X$ and are defined as 
$|\overline{0}\rangle :=
\frac{1}{\sqrt{2}}(|0\rangle +|1\rangle)$
and $|\overline{1}\rangle:=
\frac{1}{\sqrt{2}}(|0\rangle -|1\rangle)$. 
She prepares a quantum state 
of $N$ qubits described by a Hilbert space 
${\cal H}_A:={\bf C}^2 \otimes {\bf C}^2
\otimes \cdots \otimes {\bf C}^2$ ($N$ times) as follows. 
If her choice of basis is $X$,
she encodes her message $y=y_1y_2\cdots y_N \in \{0,1\}^N$
as $|\overline{y}\rangle:=
|\overline{y_1}\rangle \otimes |\overline{y_2}\rangle
\otimes \cdots \otimes |\overline{y_N}\rangle
\in {\cal H}_A$. 
If her choice of basis is $Z$, 
she encodes her message $y$ as $|y\rangle:=
|y_1\rangle \otimes |y_2\rangle \otimes \cdots \otimes |y_N\rangle
\in {\cal H}_A$. 
Alice sends thus prepared $N$ qubits to Bob. 
Eve, whose purpose is to obtain information about the message, 
makes her apparatus 
interact with the qubits sent 
from Alice to Bob and divides the whole system into 
two parts. 
This process is described by a completely-positive map (CP-map) 
\begin{eqnarray*}
\Lambda:{\cal S}({\cal H}_A) \to {\cal S}({\cal H}_B \otimes {\cal H}_E), 
\end{eqnarray*}
where ${\cal H}_B$ (resp. ${\cal H}_E$) denotes a Hilbert space 
of the system distributed to Bob (resp. Eve), and 
${\cal S}({\cal H})$ is a set of all density operators 
on a Hilbert space ${\cal H}$. 
Alice then announces the basis $X$ or $Z$ that she had used for 
encoding. 
Bob and Eve try estimating the message by 
using the quantum state and the information of the basis. 
Note that in this protocol ${\cal H}_B$ and 
${\cal H}_E$ may be general quantum systems. 
In particular, ${\cal H}_B$ may not be qubits. 
Thus in contrast to the standard quantum key distribution protocol, 
Bob may not measure $X$ or $Z$ to obtain information. 
Bob knows the basis used for encoding and 
the form of CP-map $\Lambda$. 
Thus Bob and Eve are equal in their knowledges 
on classical information. Only the distributed quantum states 
differ with each other. 
According to the information-disturbance relationship in 
Shannon's information-theoretical setting, 
if Eve's attack helps her obtain large information about the message 
encoded in the $X$ basis, Bob cannot obtain large information about 
the message encoded in the $Z$ basis. 
If the message is chosen probabilistically \cite{prob}, 
this relationship is expressed in the formula as \cite{Miyaprep}:
\begin{eqnarray*}
I(A:E|\mbox{basis} =X)+I(A:B|\mbox{basis}=Z)\leq N,
\end{eqnarray*}
where $A$ represents the random variable of the message and 
$E$ $(\mbox{resp. }B)$ represents the random variable of 
the outcome of the measurement performed by Eve (resp. Bob), 
and $I(\cdot,\cdot)$ denotes Shannon's mutual information.  
\par
We formulate the above problem 
in the algorithmic information-theoretical setting. 
Let us denote the quantum state obtained by Bob (resp. Eve)
corresponding to 
the message $z$ $(\mbox{resp. }x)$ encoded with the basis $Z$ $(
\mbox{resp. }X)$ 
by $\rho^B_z\in {\cal S}({\cal H}_B)$ $(\mbox{resp. } \sigma^E_x \in {\cal S}({\cal H}_E))$. 
That is, $\rho^B_z$ and $\sigma^E_x$ are defined by 
\begin{eqnarray*}
\rho^B_z&=&\mbox{tr}_{{\cal H}_E}(\Lambda(|z\rangle \langle z|)) \\
\sigma^E_x&=&\mbox{tr}_{{\cal H}_B}(\Lambda(|\overline{x}\rangle
\langle \overline{x}|)),
\end{eqnarray*}
where $\mbox{tr}_{{\cal H}_E}$ (resp. $\mbox{tr}_{{\cal H}_B}$) denotes 
a partial trace over ${\cal H}_E$ (resp. ${\cal H}_B$). 
Motivated by the above result in Shannon's formulation, we expect that there 
will exist some trade-off relationship between 
$K(x|\sigma^E_x,X)$ and $K(z|\rho^B_z,Z)$
\cite{identify}. 
$K(x|\sigma^E_x,X)$ is the quantum Kolmogorov complexity of the message $x$ 
encoded with $X$ 
for Eve. 
Note that Eve has quantum state $\sigma^E_x$, 
and knows $X$ (and $\Lambda$). 
$K(z|\rho^B_z,Z)$ is the quantum Kolmogorov complexity of the 
message $z$ encoded with $Z$ for Bob. 
He has quantum state $\rho^B_z$, and knows $Z$ (and $\Lambda$). 
The following is our main theorem.
\begin{theorem}
There exists a trade-off relationship for the number of 
messages that have low complexity.
For any integers $l,m \geq 0$, 
\begin{eqnarray*}
\left| 
\{z|K(z|\rho^B_z,Z)\leq l\}
\right|
+\left|
\{x|K(x|\sigma^E_x,X)\leq m\}
\right|
\leq 2^{N}\left(
1+
2^{\frac{l+m-N}{2} +c}
\right)
\end{eqnarray*}
holds, 
where $|A|$ denotes the cardinality of a set $A$ and 
$c$ is a constant depending on 
the choice of the quantum Turing machine. 
Note that the right-hand side of the above inequality gives 
a nontrivial bound for $l, m$ satisfying 
$l+m\leq N-2c$. 
\end{theorem} 
{\bf Proof:}
The proof has three parts. 
(i) An entanglement-based protocol which is related to 
the original one is introduced. 
(ii) It is shown that the number of messages that have low complexity 
can be represented by an expectation value of a certain observable 
in the entanglement-based protocol. 
(iii) The uncertainty relation is applied to show a trade-off relationship. 
\par
(i) 
Let us analyze the protocol. 
Instead of the original protocol, 
we treat an entanglement-based protocol
(E91-like protocol), 
which is related to the original one.
It runs as follows. 
Alice prepares $N$ pairs of qubits. 
She prepares each pair in the EPR state, 
$|\phi\rangle:=
\frac{1}{\sqrt{2}}
(|0\rangle \otimes |0\rangle 
+|1\rangle \otimes |1\rangle)$. 
Therefore, the whole state can be 
written as $|\phi^N\rangle
:=|\phi\rangle \otimes |\phi \rangle 
\otimes \cdots \otimes |\phi \rangle$ 
($N$ times) 
in a Hilbert space ${\cal H}_{A'} \otimes {\cal H}_{A}
$, 
where ${\cal H}_{A'} \simeq {\cal H}_A \simeq \otimes^N {\bf C}^2$. 
Alice sends qubits described by ${\cal H}_A$ to Bob. 
Before the qubits reach Bob, 
Eve makes them interact with her own apparatus, 
and divides the whole system into two parts. 
The whole dynamics is described by $(\mbox{id}_{{\cal S}({\cal H}_{A'})}
\otimes \Lambda) : {\cal S}({\cal H}_{A'}
\otimes {\cal H}_A) \to {\cal S}({\cal H}_{A'}\otimes
{\cal H}_B \otimes {\cal H}_E)$, 
where $\mbox{id}_{{\cal S}({\cal H}_{A'})}$ is an identity map 
on ${\cal S}({\cal H}_{A'})$. 
We denote by $\Theta$ the whole state 
over ${\cal H}_{A'}\otimes {\cal H}_B \otimes {\cal H}_E$ 
after this process. 
That is, it is defined by 
$\Theta=(\mbox{id}_{{\cal H}_{A'}}\otimes \Lambda)
(|\phi^N \rangle \langle \phi^N|)$. 
Alice then measures her 
qubits with the basis $X$ or $Z$, and 
announces the basis used. 
\par
It can be shown \cite{Miyaprep} that this entanglement-based 
protocol is equivalent with the original protocol 
with a probabilistically \cite{prob} chosen message.  
In fact, we can see the following correspondence. 
Define $Z_z$ for $z\in \{0,1\}^N$,  
a projection operator on ${\cal H}_{A'}$, by 
$Z_z:=|z\rangle \langle z|$. 
$\{Z_z\}$ forms a projection-valued measure (PVM). 
Probability to obtain $z\in \{0,1\}^N$ in its measurement 
is $P_Z(z):=\mbox{tr}(\Theta (Z_z\otimes {\bf 1}_B \otimes {\bf 1}_E))
=\frac{1}{2^N}$. 
In addition, a-posteriori state \cite{aposteriori} on ${\cal H}_B \otimes 
{\cal H}_E$ is calculated as 
$\Lambda (|z\rangle \langle z|)$, 
whose restriction on ${\cal H}_B$ is 
nothing but $\rho^B_z\in {\cal S}({\cal H}_B)$. 
Similarly, define $X_x$ for $x\in \{0,1\}^N$, a projection operator on 
${\cal H}_{A'}$, by 
$X_x=|\overline{x}\rangle \langle \overline{x}|$. 
It is easy to see that $\{X_x\}_{x\in \{0,1\}^N}$ forms a PVM on 
${\cal H}_{A'}$. 
For each $x\in \{0,1\}^N$, probability to obtain $x$ 
in its measurement is $P_X(x)=\frac{1}{2^N}$. 
A-posteriori state \cite{aposteriori} on ${\cal H}_B \otimes {\cal H}_E$ 
becomes $\Lambda(|\overline{x}\rangle \langle \overline{x}|)$, 
whose restriction on ${\cal H}_E$ is $\sigma^E_x$. 
\par
(ii) 
We
fix a universal quantum Turing machine $U$ and discuss 
the quantum Kolmogorov complexity with respect to it. 
Firstly let us consider 
 the complexity for Bob when the 
message $z$ is encoded with $Z$. 
Bob knows $Z$ and has a quantum system 
described by ${\cal H}_B$ whose state is $\rho^B_z$. 
This system is identified with the auxiliary input 
tape. 
That is, we investigate $K_U(z|\rho^B_z,Z)$.  
Thanks to theorem \ref{th:classical}, 
it suffices to consider only the programs 
that exactly output the message $z$ because 
the message is a classical object. 
That is, we regard 
\begin{eqnarray*}
K_{c,U}(z|\rho^B_z,Z):=\min_{q: U(q,\rho^B_z,Z)=|z\rangle}
l(q),
\end{eqnarray*}
which satisfies $K_{c,U}(z|\rho^B_z,Z)\geq K_U(z|\rho^B_z,Z)
\geq K_{c,U}(z|\rho^B_z,Z)-c'$
for some constant $c'$. 
\par
Let us denote $T_z \subset \{0,1\}^*$ a set of all programs 
that output $z$ with auxiliary inputs $\rho^B_z$ and $Z$. 
A relationship 
$K_{c,U}(z|\rho^B_z,Z)=\min_{t\in T_z} l(t)$ follows.  
Although different programs may have different halting times, 
thanks to the lemma 
proved by M\"uller (Lemma 2.3.4. in \cite{MullerPhD}),
there exists a CP-map
$\Gamma_{U,Z}: {\cal S}({\cal H}_{B} \otimes {\cal H}_I)
\to {\cal S}({\cal H}_O)$
satisfying for any $t\in T_z$ 
\begin{eqnarray*}
\Gamma_{U,Z}(\rho^B_z
 \otimes |t \rangle \langle 
 t|)=|z\rangle \langle z|,
\end{eqnarray*}
where ${\cal H}_I$ is 
a Hilbert space for 
programs, and ${\cal H}_O =\otimes^N {\bf C}_2$
is a Hilbert space 
for outputs. 
From this lemma, we obtain an important 
observation. 
If $T_z \cap T_{z'} \neq \emptyset$ holds
for some $z\neq z'$, 
$\rho^B_z$ and $\rho^B_{z'}$ are perfectly distinguishable.
In fact, as a CP-map does not increase 
the distinguishability of states, the relationships 
for $t\in T_z
\cap T_{z'}$
\begin{eqnarray*}
\Gamma_{U,Z}(\rho^B_z \otimes |t\rangle \langle t|)
=|z\rangle \langle z| 
\\
\Gamma_{U,Z}(\rho^B_{z'} \otimes |t\rangle \langle t|)
=|z' \rangle \langle z'| 
\end{eqnarray*}
and their distinguishability on the right-hand sides 
imply the distinguishability of 
$\rho^B_z$ and $\rho^B_{z'}$. 
For each $t\in \{0,1\}^*$
we define ${\cal C}_t \subset \{0,1\}^N$ 
as ${\cal C}_t=\{z| t\in T_z\}$. 
That is, $z\in {\cal C}_t$ is a message 
which can be reconstructed by giving 
a program $t$ to the Turing machine $U$ with 
an auxiliary input $\rho^B_z$ and $Z$.  
Owing to the distinguishability between $\rho^B_z$ 
and $\rho^B_{z'}$, for $z,z'\in {\cal C}_t$, 
there exists a family of projection operators 
$\{E^t_z\}_{z\in {\cal C}_t}$ on ${\cal H}_B$ satisfying
for any $z,z' \in {\cal C}_t$, 
\begin{eqnarray*}
&& E^t_z E^t_{z'} =\delta_{z z'} E^t_z
\\
&& \sum_{z\in {\cal C}_t}E^t_z \leq {\bf 1}
\\
&& \mbox{tr}(\rho^B_z E^t_{z'})=\delta_{zz'}.
\end{eqnarray*}
As we are interested in minimum length 
programs, we define 
${\cal D}_t:=\{z| t=\mbox{argmin}_{s\in T_z} l(s)\}$, 
which is a subset of ${\cal C}_t$. 
$z\in {\cal D}_t$ is a message that  
has 
$t$ as its minimum length program for reconstruction.
It is still possible that ${\cal D}_t \cap {\cal D}_{t'}
\neq \emptyset$. That is, there may be a message $z$ 
whose shortest 
programs are not unique. 
In such a case, we choose one of the programs to 
avoid counting doubly. 
For instance, this can be done by introducing 
a total order $<$ in all the programs $\{0,1\}^*$, 
and by defining 
${\cal E}_t=\{z|z\in {\cal D}_t, z \notin {\cal D}_{t'}
\mbox{ for all }t'<t\mbox{ with }l(t)=l(t')\}$. 
As this ${\cal E}_t$ is a subset of 
${\cal C}_t$, for any $z, z'\in {\cal E}_t$ 
\begin{eqnarray}
&& E^t_z E^t_{z'} =\delta_{z z'} E^t_z
\nonumber \\
&& \sum_{z\in {\cal E}_t}E^t_z \leq {\bf 1}
\nonumber 
\\
&& \mbox{tr}(\rho^B_z E^t_{z'})=\delta_{zz'}
\nonumber
\end{eqnarray}
hold. 
\par 
For any program $t\in \{ 0,1\}^*$ 
we define a projection operator
$P_t:=\sum_{z\in {\cal E}_t}
(Z_z \otimes E^t_z \otimes {\bf 1}_E)$. 
For any integer $l\geq 0$, 
we consider a projection operator 
$\hat{P}_l:=\sum_{t: l(t)\leq l}
P_t$, 
whose 
expectation value 
with $\Theta$ becomes
\begin{eqnarray}
\mbox{tr}(\Theta \hat{P}_l)
&=&
\sum_{t: l(t)\leq l}
\sum_{z\in {\cal E}_t}
P_Z(z)\mbox{tr}(\rho^B_z E^t_z)
\nonumber \\
&=&\sum_{t: l(t)\leq l}\sum_{z\in {\cal E}_t} P_Z(z)
\nonumber \\
&=&\frac{1}{2^N} 
\left|
\{z|K_{c,U}(z|\rho^B_z,Z)\leq l\}
\right|
.\label{Pl}
\end{eqnarray}
\par
Similarly, we treat 
$K_{c,U}(x|\sigma^E_x,X)$. 
We can introduce 
$S_x \subset \{0,1\}^*$ a set of all programs 
that output $x$ with auxiliary inputs $\sigma^E_x$ and $X$. 
$K_{c,U}(x|\sigma^E_x,X)=\min_{s\in S_x} l(s)$ holds. 
We can define ${\cal J}_s:=\{x|s \in S_x\}$ for each $s$
and introduce a family of 
projection operators $\{F^s_x\}_{x\in {\cal F}_s}$ 
on ${\cal H}_E$ 
that satisfies 
\begin{eqnarray*}
\mbox{tr}(F^s_x \sigma^E_{x'})=\delta_{xx'}
\end{eqnarray*} 
for each $x,x'\in {\cal J}_s$ and so on.
${\cal G}_s:=\{x| s=\mbox{argmin}_{t\in S_x} l(t) \}$
and 
${\cal F}_s:=\{z|z\in {\cal G}_s, z \notin {\cal G}_{s'}
\mbox{ for all }s'<s\mbox{ with }l(s)=l(s')\}$, 
are also defined. 
We consider a family of projection operators
$\{F^s_x \}_{x\in {\cal F}_s}$.
Similarly, 
for any program $s\in \{0,1\}^*$, 
we define a projection 
operator $Q_s :=
\sum_{x\in {\cal F}_s}
(X_x \otimes {\bf 1}_B \otimes F^s_x)$ 
and consider 
for any integer $m\geq 0$, 
$\hat{Q}_m:=\sum_{s: l(s)\leq m}
Q_s$, 
whose expectation value with respect to $\Theta$ 
is written as
\begin{eqnarray}
\mbox{tr}(\Theta \hat{Q}_m)
=\frac{1}{2^N}
\left|
\{x|K_{c,U}(x|\sigma^E_x,X)\leq m\}
\right|
. \label{Qm}
\end{eqnarray}
\par
(iii) 
Our purpose is to obtain a trade-off 
relationship between (\ref{Pl}) and (\ref{Qm}).
It is obtained by applying 
the uncertainty relation, which 
is often regarded as the most 
fundamental inequality characterizing 
quantum mechanics. 
Among the various forms of 
the uncertainty relation, 
we employ the Landau-Pollak uncertainty relation
for arbitrary numbers of projection operators \cite{MiyaLP}. 
For a finite family of projection operators $\{A_i\}$
and any state $\rho$,  
it holds that 
\begin{eqnarray*}
\sum_i \mbox{tr}(\rho A_i)
\leq 1 +
\left( \sum_{i\neq j}
\Vert A_i A_j\Vert^{2} 
\right)^{1/2}. 
\end{eqnarray*}
We apply this inequality for 
a family of projection operators
$\{P_t, Q_s\}\ (l(t)\leq l, l(s)\leq m)$ 
and the state $\Theta$. 
As $P_t P_{t'}=0$ for $t\neq t'$ and 
$Q_s Q_{s'} =0$ for $s\neq s'$ hold
thanks to ${\cal E}_t \cap {\cal E}_{t'}
={\cal F}_s \cap {\cal F}_{s'}
=\emptyset$, 
we obtain 
\begin{eqnarray*}
\mbox{tr}(\Theta \hat{P}_l)
+\mbox{tr}(\Theta \hat{Q}_m)
\leq 1
+\left(2
\sum_{t}^{l(t)\leq l} \sum_{s}^{l(s)\leq m}
\Vert P_t Q_s\Vert^2
\right)^{1/2}. 
\end{eqnarray*}
The term $\Vert P_t Q_s\Vert$ of the right-hand side 
is 
computed as follows. 
As the operator norm $\Vert P_t Q_s\Vert$ is 
written as $\Vert P_t Q_s \Vert 
=\sup_{|\Psi\rangle:\Vert | \Psi\rangle \Vert=1}
\Vert P_t Q_s |\Psi\rangle \Vert$, 
we need to bound $\Vert  P_t Q_s
|\Psi\rangle \Vert$ for any normalized 
vector $|\Psi\rangle$.
\begin{eqnarray*}
\Vert \sum_{z\in {\cal E}_t}
\sum_{x\in {\cal F}_s}
(Z_z X_x \otimes E^t_z \otimes F^s_x) |\Psi\rangle \Vert
&=&
\left(
\sum_{z\in {\cal E}_t}
\sum_{x\in {\cal F}_s}
\langle \Psi |
(X_x Z_z X_x \otimes E^t_z \otimes F^s_x) |\Psi\rangle
\right)^{1/2}
\\
&=&
\left(
\sum_{z\in {\cal E}_t}
\sum_{x\in {\cal F}_s}
\mbox{tr}(\mu^{t,s}_{z,x}X_x Z_z X_x)
\langle \Psi |{\bf 1}_A \otimes E^t_z
\otimes F^s_x|\Psi\rangle
\right)^{1/2}
,
\end{eqnarray*}
where we used $E^t_z E^t_{z'} =0$ for $z\neq z'$ and 
$F^s_x F^s_{x'}=0$ for $x\neq x'$, 
and 
$\mu^{t,s}_{z,x}$ is a-posteriori state \cite{aposteriori} 
defined 
as a unique state 
satisfying the above equality. 
\\
As $
|\mbox{tr}(\mu^{t,s}_{z,x}X_x Z_z X_x)|
\leq \Vert X_x Z_z X_x\Vert 
=\frac{1}{2^N}$ holds, 
we obtain
\begin{eqnarray*}
\left(
\sum_{z\in {\cal E}_t}
\sum_{x\in {\cal F}_s}
\mbox{tr}(\mu^{t,s}_{z,x}X_x Z_z X_x)
\langle \Psi |{\bf 1}_A \otimes E^t_z
\otimes F^s_x|\Psi\rangle
\right)^{1/2}
&\leq&
\frac{1}{2^{N/2}}
\left(
\sum_{z\in {\cal E}_t}
\sum_{x\in {\cal F}_s}
\langle \Psi |{\bf 1}_A \otimes E^t_z
\otimes F^s_x|\Psi\rangle
\right)^{1/2}
\\
&\leq& \frac{1}{2^{N/2}}, 
\end{eqnarray*}
where we have used 
$\sum_{z\in {\cal E}_t} E^t_z \leq {\bf 1}_B$ and 
$\sum_{x\in {\cal F}_s} F^s_x \leq {\bf 1}_E$. 
As $|\{t|l(t)\leq l\}|\leq 2^{l+1}$
and $|\{s|l(s)\leq m\}|\leq 2^{m+1}$ 
hold, 
we obtain
\begin{eqnarray*}
\mbox{tr}(\Theta \hat{P}_l)
+\mbox{tr}(\Theta \hat{Q}_m)
\leq 1
+2^{\frac{l+m-N+3}{2}}.
\end{eqnarray*}
This inequality with 
(\ref{Pl}) and (\ref{Qm}) derives
\begin{eqnarray*}
\left| 
\{z|K_{c,U}(z|\rho^B_z,Z)\leq l\}
\right|
+\left|
\{x|K_{c,U}(x|\sigma^E_x,X)\leq m\}
\right|
\leq 2^{N}\left(
1+
2^{\frac{l+m-N+3}{2}}
\right). 
\end{eqnarray*}
Taking into consideration the relationship between $K_{c,U}$ 
and $K_U$, we finally obtain 
\begin{eqnarray*}
\left| 
\{z|K(z|\rho^B_z,Z)\leq l\}
\right|
+\left|
\{x|K(x|\sigma^E_x,X)\leq m\}
\right|
\leq 2^{N}\left(
1+
2^{\frac{l+m-N}{2} +c}
\right),
\end{eqnarray*}
where $c$ is a constant.
\hfill Q.E.D.
\par
Let us consider the implication of the above theorem. 
As noted in the theorem, a nontrivial bound is given 
only for $l+m \leq N-2c$. This situation is attained 
when one considers the asymptotic behavior of 
a family of protocols 
governed by increasing $N$. 
We consider $\{z| K(z|\rho^B_z,Z)\leq p_Z N\}$ and 
$\{x| K(x|\sigma^E_x,X)\leq p_xN\}$ for some 
$p_Z, p_X\in [0,1]$. If $p_Z$ and $p_X$ satisfy 
$p_Z +p_X <1$, for a sufficiently large $N>0$,  
the right-hand side of the above theorem behaves as  
$2^N(1+O(2^{-\epsilon N}))$ for some $\epsilon>0$. 
That is, for any $p_Z, p_X\in [0,1)$ satisfying 
$p_X+p_Z<1$, there exists $\epsilon>0$ such that it holds 
\begin{eqnarray*}
|\{z|K(z|\rho^B_z,Z)\leq p_Z N\}|
+|\{x|K(x|\sigma^E_x,X)\leq p_X N\}|
\leq 2^N (1+O(2^{-\epsilon N})). 
\end{eqnarray*}
This type of argument is common in the algorithmic 
information theory. 
\par
In addition, the above theorem gives the following corollaries, 
which should be meaningful for an asymptotically 
large $N$.  
\begin{corollary}\label{cor1}
There exists a trade-off relationship between 
$\max_z
K(z|\rho^B_z,Z)$ and 
$\max_x
K(x|\sigma^E_x,X)$:  
\begin{eqnarray*}
\max_{z\in \{0,1\}^N} K(z|\rho^B_z,Z)+\max_{x\in \{0,1\}^N} K(x|\sigma^E_x,X)
\geq N-O(1).
\end{eqnarray*}
\end{corollary}
{\bf Proof:}
Because for $l=\max_z K(z|\rho^B_z,Z)$ and 
$m=\max_x K(x|\sigma^E_x,X)$, $\{z|K(z|\rho^B_z,Z)\leq l\}|
=2^N$ and $\{x|K(x|\sigma^E_x,X)\leq m\}|=2^N$ hold, 
the right-hand side of the above theorem must be larger than 
$2^N(1+1)$.  It is only possible when 
$l+m\geq N-2c$ holds. 
\hfill Q.E.D.
\begin{corollary}(No-cloning theorem 
\cite{WZ,MiyaLogics})
Unknown states cannot be cloned (for a sufficiently large $N$).
\end{corollary}
{\bf Proof:}
Suppose that universal cloning is possible. 
Put ${\cal H}_B\simeq {\cal H}_E \simeq {\cal H}_A$. 
There should exist a CP-map $\Lambda$ satisfying both 
$\Lambda(|z\rangle \langle z|)=|z\rangle \langle z|
\otimes |z\rangle \langle z|$ and 
$\Lambda(|\overline{x}\rangle \langle \overline{x}|)
=|\overline{x}\rangle \langle \overline{x}|
\otimes |\overline{x}\rangle \langle \overline{x}|$ 
for all $z,x\in \{0,1\}^N$. 
It implies that $\max_z K(z|\rho^B_z,Z)=O(1)$
and $\max_x K(x|\sigma^E_x,X)=O(1)$. 
This contradicts corollary \ref{cor1}. 
\hfill Q.E.D.
\section{Discussion}
In this research, we study a quantum algorithmic 
information-theoretic representation of the 
information-disturbance theorem.  
We first discuss the relationship between 
Shannon's information-theoretic theorem and 
our algorithmic one. 
Using a possible relationship between 
Shannon information and Kolmogorov complexity 
is likely to yield an inequality 
\begin{eqnarray}
\sum_{z\in \{0,1\}^N}p_Z(z)
K(z|\rho^B_z,Z)+\sum_{x\in\{0,1\}^N}p_X(x)
K(x|\sigma^E_x,X)\geq N-c,
\label{tabun}
\end{eqnarray}
directly from Shannon's version. 
This inequality is different from our theorem derived 
in the present paper. In fact, even if families $\{K(z|\rho^B_z,Z)\}_z$ 
and $\{K(x|\sigma^E_x,X)\}_{x}$ satisfy this inequality, 
they may not satisfy the inequality in our theorem. 
In fact, if we put 
$|\{z|K (z|\rho^B_z,Z)=\frac{N}{2}\}|=\frac{3\cdot 2^N}{4}$,
$|\{z|K(z|\rho^B_z,Z)=N\}|=\frac{2^N}{4}$, 
$|\{x|K(x|\sigma^E_x,X)=\frac{N}{3}\}|=\frac{3\cdot 2^N}{4}$,
and $|\{x|K(x|\sigma^E_x,X)=N\}|=\frac{2^N}{4}$, 
then 
the left-hand side of (\ref{tabun}) becomes $\frac{9N}{8}$, 
but our theorem (with $c=0$) is not satisfied for $l=\frac{N}{2}$ and 
$m=\frac{N}{3}$.  
It would be interesting to investigate 
an inequality for Shannon's information that 
corresponds to our theorem. 
\par
As mentioned in the introduction, one of the purposes of this study is 
to demonstrate the usage of quantum Kolmogorov complexity 
in the quantum information theory. Our derivation dealt with  
Kolmogorov complexity directly without relying on 
the results known in Shannon's version of the quantum information theory. 
Those results imply that Kolmogorov complexity can yield 
meaningful results by combining it with 
the uncertainty relation. 
Thus, quantum Kolmogorov complexity by itself can be a powerful tool 
in the quantum information theory by itself.
\par
In addition, as mentioned earlier, there are various quantum versions 
of Kolmogorov complexity.  It would be interesting and important to 
study quantum information-theoretic problems by using these quantum versions. 
\par 
While our information-disturbance theorem was formulated 
in a cryptographic setting, 
it is strongly related to 
Heisenberg's uncertainty principle, which is one of the 
most important characteristics of quantum mechanics. 
According to Heisenberg's original 
Gedanken experiment, a 
precise measurement of the momentum
destroys the position of a particle. 
If one regards Eve's attack as 
the measurement of ``momentum",
the information-disturbance relationship 
that predicts a disturbance in the conjugate ``position" 
corresponds to the 
Heisenberg's setting. 
However, despite this similarity, there is a gap between 
our information-disturbance theorem and 
Heisenberg's uncertainty principle. 
The latter should be formulated as a
relationship that does not depend on states 
as was discussed in \cite{MiyaHeisen}. 
Further investigation in this direction 
needs to be carried out. 
Besides exploring subjects related to 
the uncertainty principle, 
several things need to be done. 
We hope that the quantum Kolmogorov complexity will shed 
new light on the quantum information theory. 

\section*{Acknowledgments}
I would like to thank K.Imafuku and anonymous referees for valuable discussions and comments.






\bibliographystyle{mdpi}

\makeatletter

\renewcommand\@biblabel[1]{#1. }

\makeatother




\begin{thebibliography}{1}
\bibitem{prob}The word ``probabilistically" here is used to 
mean ``randomly in a probabilistic sense". 
That is, 
we use an unbiased probability $1/|\Omega|$ to choose a sample 
from a sample space $\Omega$ (say $\Omega=\{0,1\}^{2N}$).
(To avoid a 
possible confusion of it with randomness in 
algorithmic sense, we just write ``probabilistically".)
\bibitem{identify}
To treat $\rho^B_z$ and $\sigma^E_x$ as an auxiliary input
for a quantum Turing machine, they have to be somehow
represented as states on a system consisting of qubits. 
Our discussion does not depend on how we identify them.
\end{thebibliography}


\end{document}